\documentstyle[12pt]{article}

\begin{document}
\input{epsf.sty}
\baselineskip 15pt

\title{
ARE INTERACTION-FREE MEASUREMENTS INTERACTION FREE?}
\author{ Lev Vaidman}
\maketitle

\begin{center}
{\small \em School of Physics and Astronomy \\
Raymond and Beverly Sackler Faculty of Exact Sciences \\
Tel Aviv University, Tel-Aviv 69978, Israel. \\}
\end{center}

\vspace{.3cm}
\begin{abstract}
{\small
In 1993 Elitzur and Vaidman introduced the concept of interaction-free
measurements which allowed  finding objects without ``touching'' them.
In the proposed method, since the objects were not touched even by photons,
thus, the interaction-free measurements can be called as ``seeing in the
dark''. Since then several experiments have been successfully performed and
various modifications were suggested. Recently, however, the validity of the
term ``interaction-free'' has been questioned. The criticism of the name is
briefly reviewed and the meaning of the interaction-free measurements is
clarified.}
\end{abstract}

\vskip 0.5cm \noindent
{\bf  1. Introduction ~~}
\vskip .1cm

The interaction-free measurements proposed by Elitzur and Vaidman
\cite{EV91,EV93} (EV IFM) led to numerous investigations and several
experiments have been performed
\cite{Kwi95,Voo97,HaSu97,Tse98,White98,Kwi98,MiMi99}.  While there is a
consensus about importance of this proposal, there have been several
objections to the name ``interaction-free''. Some authors in trying
to avoid it, made modifications such as ``interaction (energy
exchange) free measurements'' \cite{en-ex1,en-ex2}, ``indirect
measurements'' \cite{indi}, ``seemingly interaction-free
measurements'' \cite{seem-ifm}, ``interaction-free'' interrogation
\cite{Kwi98}, etc.  Moreover, recently, Simon and Platzman
\cite{SiPl} claimed that there is a ``fundamental limit on
`interaction-free' measurements''.

The discussion of the term ``interaction-free'' appears in the
original IFM paper \cite{EV93}, but reading papers about the
interaction-free measurements has convinced me that the concept of EV
IFM has been 
frequently misunderstood.  In this paper I want to clarify in which
sense the interaction-free measurements are interaction free.  I will
also make a comparison with procedures termed ``interaction-free
measurements'' in the past and will analyze conceptual advantages and
disadvantages of various modern schemes for the interaction-free
measurements.

\vskip 0.5 cm \noindent
{\bf  2. How the EV IFM paper was written? ~~}
\vskip .1cm

At the beginning of 1991, Avshalom Elitzur came to me with the
following question: {\it Suppose there is an object such that {\em any}
  interaction with it leads to an explosion. Can we locate the object
  without exploding it?} Our joint work resulted in a positive answer
to this question described in the EV IFM paper.

Presented in this way, the name interaction-free is clearly
appropriate. Simple logic tells us: given that any interaction leads
to an explosion and given that there has been no explosion, it follows
that there have been no interaction. This way of reasoning was
described in Section 4 of our paper.  However, the proposed method
have certain additional features which justify the name
``interaction-free''. The method is applicable for location of objects
which do not necessarily explode. Even for such objects we can claim
that, in some sense, the finding of its location is
``interaction-free''. These aspects of interaction-free measurements
were explained at the beginning of the EV IFM paper.  Before I 
continue with the discussion let me briefly describe our solution to
the posed question.

\vskip 0.5cm \noindent
{\bf  3. The original proposal for the IFM.  ~~}
\vskip .1cm

Our method is based on the Mach-Zehnder interferometer.  A photon
(from a source of single photons) reaches the first beam splitter
which has transmission coefficient ${1\over2}$.  The transmitted and
reflected parts of the photon's wave are then reflected by the mirrors
and finally reunite at another, similar beam splitter (Fig.~1{\it a}).
Two detectors are positioned to detect the photon after it passes
through the second beam splitter.  We arrange the positions of the
beam splitters and the mirrors such that (because of destructive
interference) the photon is never detected by one of the detectors,
say $D_2$, and is always detected by $D_1$.  We next position the
interferometer in such a way that one of the routes of the photon
passes through the place where the object (a bomb) might be present
(Fig.~1{\it b}).  We send a single photon through the system.  There
are three possible outcomes of this measurement:\hfil \break
\phantom{.}~~~~ i)~explosion,~~~ ii)~detector $D_1$ clicks,~~~
iii)~detector $D_2$ clicks.
 
\noindent
If the detector $D_2$ clicks
(the probability for that is ${1\over4}$), we have achieved our goal:
we know that  the object  is inside the interferometer
and it did not explode.
 
\vskip 0.5 cm \noindent
{\bf 4. Measurement without ``touching''~~.}

\vskip .1cm

 In the IFM paper, we have claimed that, in some sense,
 we locate the object without touching it. However, we wrote 

 \begin{quotation}
   The argument which claims that this is an interaction-free
   measurement sounds very persuasive but is, in fact, an artifact of
   a certain interpretation of quantum mechanics (the interpretation
   that is usually adopted in discussions of Wheeler's delayed-choice
   experiment). The paradox of obtaining information without
   interaction appears due to the assumption that only one ``branch''
   of a quantum state exists. (p. 991)
 \end{quotation}

\epsfysize=8.3 cm 
\centerline{ \epsfbox{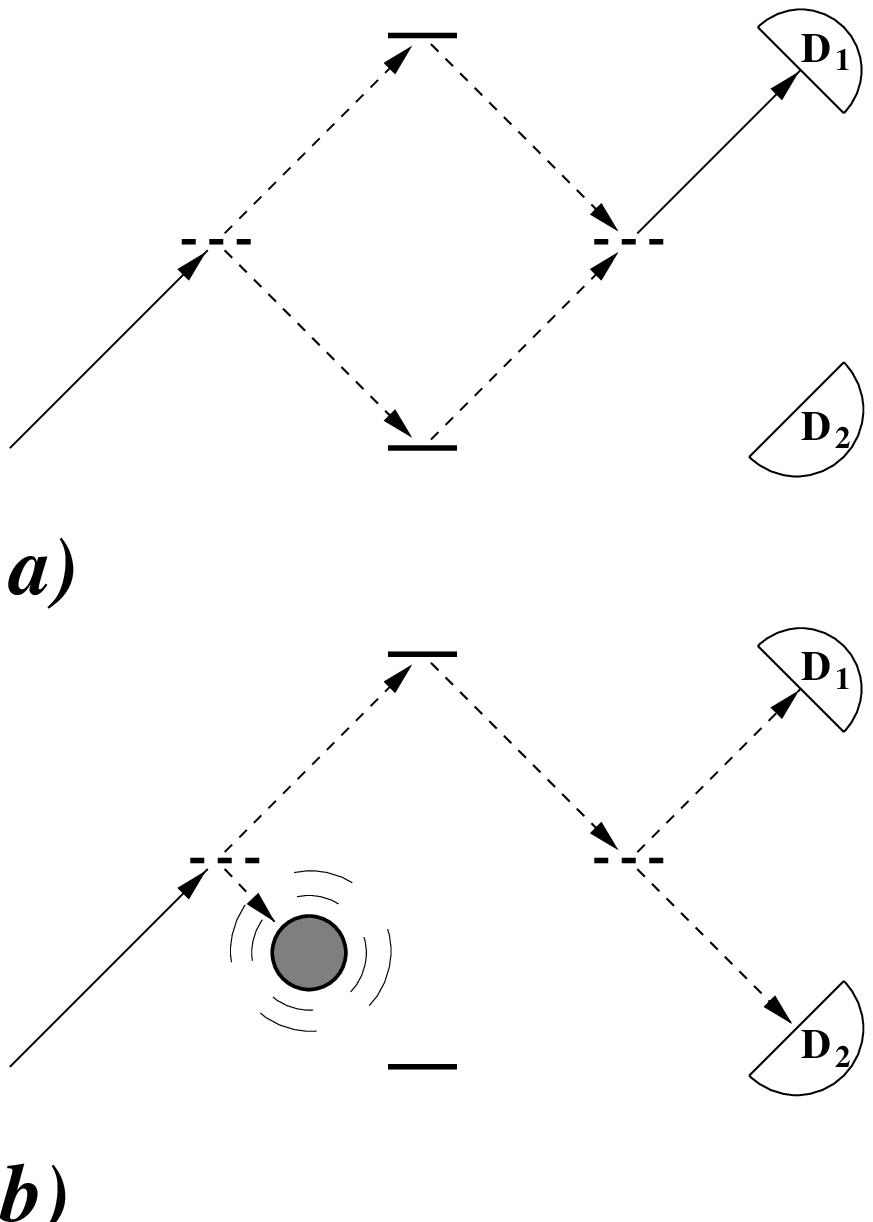}}

\noindent 
{\small {\bf Figure 1.~~}(a) When  the interferometer is
  properly tuned, all photons are detected by $D_1$ and none reach
  $D_2$. The mirrors must be massive enough and have well-defined
  position.\hfill\break (b) If the bomb is present, detector $D_2$ has
  probability 25\% to detect the photon we send through the
  interferometer and in this case we know that the bomb is inside the
  interferometer without exploding it.  }

\epsfysize=8.3 cm
 \centerline{\epsfbox{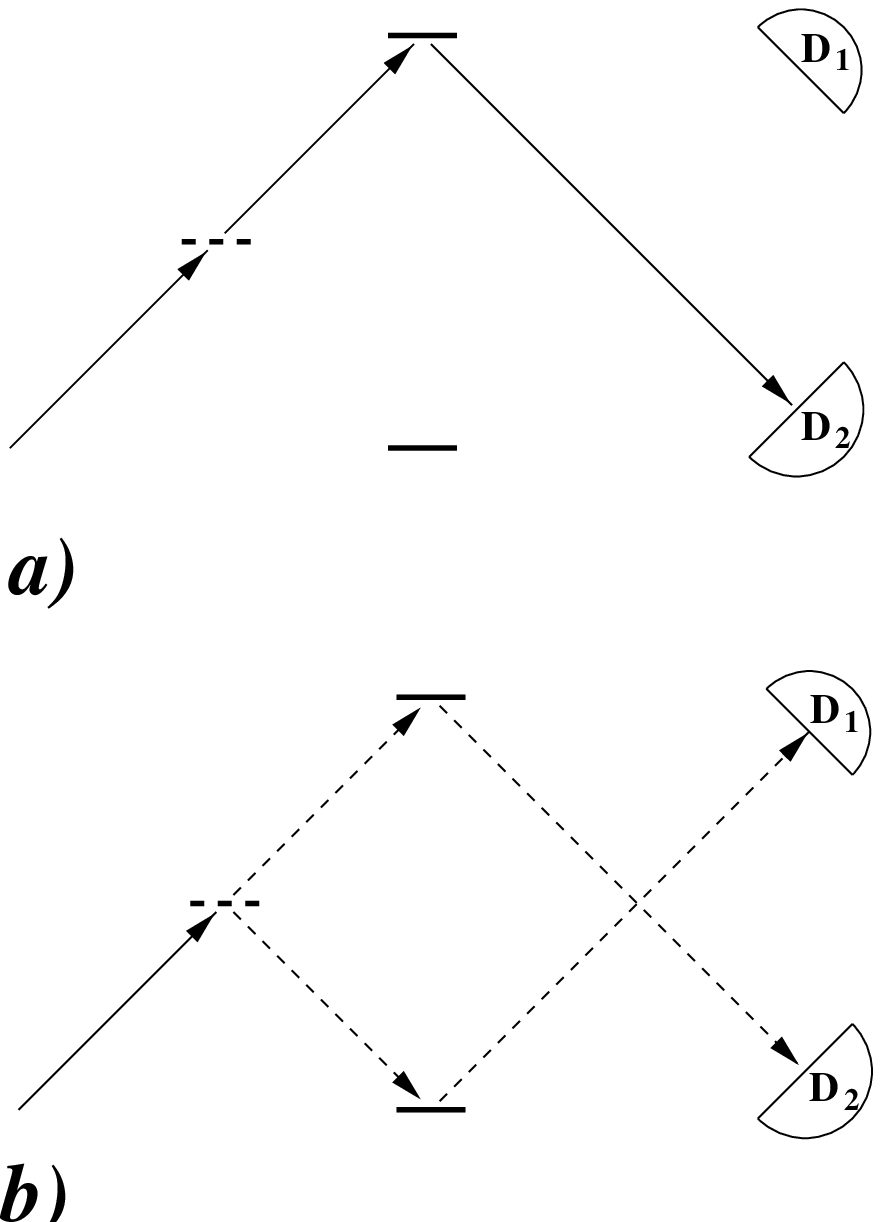}} 

\noindent 
 {\small {\bf Figure 2.~~} (a) The ``trajectory'' of the photon in the
   Wheeler experiment given that $D_2$ detected the photon as it
   usually described. The photon cannot leave any physical trace
   outside its ``trajectory''.  \hfill\break (b) The ``trajectory''
   of the quantum wave of the photon in the Wheeler experiment
   according to the von Neumann approach. The photon remains in a
   superposition until the collapse which takes place when one of the
   wave packets reaches a detector.  }

 One of the ``choices'' of Wheeler's delayed-choice experiment is an
 experiment with a Mach-Zehnder interferometer in which the second beam
 splitter is missing (see Fig.~2). In the run of the experiment with a
 single photon detected by $D_2$, it is usually accepted that the
 photon had a well defined trajectory: the upper arm of the
 interferometer. In contrast, according to the von Neumann approach,
 the photon was in a superposition inside the interferometer until the
 time when one part of superposition reached the detector $D_2$ (or
 until the time the other part reached the detector $D_1$ if that
 event was earlier). At that moment the wave function of the photon
 collapses to the vicinity of $D_2$.  The justification of Wheeler's
 claim that the photon detected by $D_2$ never was in the lower arm of
 the interferometer is that, according to the quantum mechanical laws,
 we cannot see any physical trace from the photon in the lower arm of
 the interferometer. This is true if (as it happened to be in this
 experiment) the photon from the lower arm of the interferometer
 cannot reach the detector $D_2$. The fact that there cannot be
 physical trace of the photon in the lower arm of the interferometer
 can be explained in the framework of the two-state vector formulation
 of quantum mechanics \cite{ABL,AV90}.  This formalism is particularly
 suitable for this case because we have pre- and post-selected
 situation: the photon was post-selected at $D_2$. Thus, while the wave
 function of the photon evolving forward in time does not vanish in the
 lower arm of the interferometer, the backward-evolving wave function
 does.  Vanishing one of the waves (forward or backward) at a
 particular location is enough to ensure that the photon cannot cause
 any change in the local variables of the lower arm of the
 interferometer.

 In our experiment we have the same situation. If there is an object
 in the lower arm of the interferometer, the photon cannot go through
 this arm to the detector $D_1$. This is correct if the object is such
 that it explodes whenever the photon reaches its location and we have
 not observed the explosion. Moreover, this is also correct in the
 case in which the object is completely not transparent and it blocks
 the photon in the lower arm eliminating any possibility of reaching
 $D_1$.  Even in this case we can claim that we locate the object
 ``without touching''. This claim is identical to the argument
 according to which the photon in Wheeler's experiment went solely
 through the upper arm.  In the framework of the two-state vector
 approach we can say that the forward-evolving quantum state is
 nonzero in the lower arm of the interferometer up to the location of
 the object, while the backward-evolving wave function is nonzero from
 the location of the object. Thus, at every point of the lower arm of
 the interferometer one of the quantum states vanishes. This ensures
 that the photon cannot make any physical trace there. Note, that the
 two-state vector formalism itself does not suggest that the photon is
 not present at the lower arm of the interferometer; it only helps to
 establish that the photon does not leave a trace there. The latter is
 the basis for the statement that in some sense the photon was not
 there.

\vskip 0.5cm \noindent
{\bf  5. Nested interaction-free measurements. ~~}
\vskip .1cm

   There is a very puzzling point regarding interaction-free localization of a
   quantum object which can itself be in a superposition of being in
   different locations. Our method works well for this case too (see
   Sec. 3 of the IFM paper). If $D_2$ clicks, the object is localized
   inside the interferometer. If we assume that before the experiment
   the whole volume of the interferometer except the ``working area''
   which we want to test was found empty, we can claim that the click
   of $D_2$ localizes the object inside the ``working area''. We can
   make this claim safely because we are sure that any test of our
   statement will invariably show that we are right. The object (if
   observed) will be found in the ``working area'' with certainty.
   
   However, surprisingly, the click of $D_2$ is not enough to claim
   that the photon was not in the lower arm. Indeed, the object could
   be itself a ``particle'' of another interaction-free measurement
   (we can consider a gedanken situation in which the object which
   explodes when the photon reaches its location can, nevertheless, be
   manipulated by other means). If the latter was successful (i.e. its
   ``$D_1$'' clicked) the other observer can claim that he localized
   the single photon of the first experiment in the ``working area'',
   i.e. that the photon passed through the lower arm of the
   interferometer on its way to $D_2$ \cite{Hardy}. Paradoxically, both
   claims are true: the first experiment localizes the object in the
   working area, and the second, at the same time, localizes the
   single photon there.  Both claims are true separately, but not
   together: if we would try to locate both the photon and the object
   in the working area, we will fail with certainty. Such
   peculiarities take place because we consider a  pre- and post-selected
   situation (the post-selection is that in both experiments detectors
   $D_1$ click) \cite{Har-Vai}.  In spite of this peculiar feature,
   the experiment is still interaction-free in the above sense: if we
   locate an object in a particular place, we can claim that no
   photon was at the vicinity of this place.

\vskip 0.5cm \noindent
{\bf  6. The IFM of Renninger and Dicke and the  EV IFM. ~~}
\vskip .1cm

In many papers describing experiments and modifications of the EV IFM
the first cited papers are one by Renninger \cite{Renn} and another by
Dicke \cite{Dick}. It is frequently claimed that Elitzur and Vaidman
``extended ideas of Renninger and Dicke'' and Geszti \cite{Gesz} even
wrote that we just ``amplified the argument by inventing an efficient
interferometric set''.  In fact, we came to the idea of the IFM
without any connection to these papers (the first was translated to me
only recently). We do cite Dickes' paper, although, what we got from it is
just the name: ``interaction-free measurements'' but not the method, and,
more importantly, Dickes' paper does not address the question we have solved. It seems to me that
there is very little in common between Renninger-Dicke  IFM and
EV IFM.

Renninger considered a spherical wave of a photon after it extended
beyond the radius at which a scintillate detector was located in the
part of the space angle (see Fig.~3). He discussed a {\it negative
  result experiment}: a situation in which the detector does not
detect anything. The state of the detector remained unchanged but,
nevertheless, the wave-function of the photon is modified. Dicke
considered an atom in a ground state inside a potential well. Half of
the well was illuminated by a beam of photons. Again, a negative
result experiment was considered in which no scattered photons were
observed. Dicke concentrated on the question of conservation of energy
in this experiment. The atom changed its state from the ground state
to some superposition in which the atom does not occupy the half of
the well illuminated by the photons, while photons did not change
their state at all, and he asked: ``What is the source of additional
energy of the atom?!''

The word ``measurement'' in quantum theory have many very different
meanings \cite{Bell}.  The purpose of the Renninger and Dicke measurements
is {\it preparation} of a quantum state. In contrast, the purpose of
the EV interaction-free measurement is to obtain {\it information}
about the object. In Renninger and Dicke measurements the {\it
  measuring device} is undisturbed (these are negative result
experiments) while in the EV measurement the {\it observed object} is,
in some sense, undisturbed.  In fact, in general EV IFM the quantum
state of the observed object {\it is} disturbed: the wave function
becomes localized at the vicinity of the lower arm of the
interferometer (see Sec. 3 of the EV paper). The reasons for using the
term ``interaction-free measurements'' are that the object does not
explode (if it is a bomb), it does not absorb any photon (if it is an
opaque object) and that we can claim that, in some sense,  the photon
does not reach the vicinity of the object.

A variation of Dicke's measurement which can serve as the measurement
of location of an object was considered in the original paper for
justifying the name interaction-free measurements of the EV procedure.
An object in a superposition of being in two far away places was
considered. A beam of light passed through one of the locations and no
scattered photons were observed. We obtain information that the object
is located in the other place.  This experiment is interaction-free
because the object (if it is a bomb) would not explode: the object is
found in the place where there were no photons. In such an experiment,
however, it is more difficult to claim that the photon was not at the
vicinity of the object: the photon was not at the vicinity of the {\it
  future} location of the object. But the main weakness of this
experiment relative to the EV scheme is that we get information about
the location of the object only if we have {\it prior information} about
the state of the object. If it is known in advance that the object can
be found in one out of two boxes and it was not found in one,
obviously, we then know that it is in the second box.  The whole
strength of the EV method is that we get information that an object is
inside the box {\it without any prior information!}  The latter,
contrary to the former task cannot be done without help of a quantum
theory.

Note that Dicke named his experiment ``interaction-free'' because of
another reason: the photons did not scatter: this is a ``negative
result experiment''. By using the same term we, ourselves, caused this
confusion.

 \vskip .3cm

\epsfysize=5.3 cm
 \centerline{\epsfbox{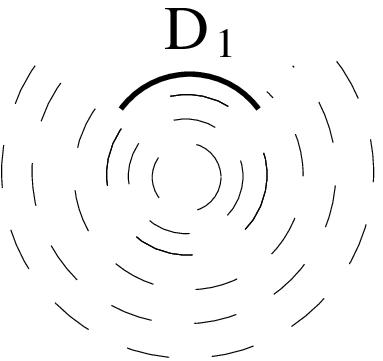}} 

\vskip .1cm
\noindent
{\small {\bf Figure 3.~ Renninger experiment} The photon spherical wave is
  modified by  the scintillate detector $D_1$ in spite of the fact that
  it detected nothing.}

\break

\vskip 0.5cm \noindent
{\bf  7. The non-demolition measurements, energy-exchange free, momentum-exchange free
  measurements and the  EV IFM.~~}
\vskip .1cm

A basic concept of quantum measurement theory is a {\it
  non-demolition measurement}. A non-demolition measurement of a
variable $A$ leaves the quantum state of the object undisturbed
provided it was in an eigenstate of $A$ prior to the measurement. It is
not an easy task to perform a non-demolition measurement.\cite{Brag}
The EV method can be applied for performing various non-demolition
measurements\cite{en-ex2}. Indeed, even if the measurement interaction
can
destroy the object, the method also  allows measurement without disturbing
the object.  The EV interaction-free measurement is a non-demolition
measurement of the position of an object (more precisely, the
measurement of the projection operator on a certain region). However,
not {\it any} non-demolition measurement of the position of the object
is an interaction-free measurement. There are methods of
non-demolition measurements in which, in the process of measurement
the state of the object changes, but these changes are compensated at
the end of the process \cite{Aha-comp}. Such measurements should not
be considered as interaction-free measurements.

Probably, the largest misconception about the IFM is defining them as
momentum-exchange free measurements \cite{SiPl}. The EV IFM can
localize a bomb in an arbitrary small region without exploding it even
if the quantum state of the bomb was spread out initially.
Localization of an object without uncertain change in its momentum
leads to immediate contradiction with the Heisenberg uncertainty
principle. Identifying the interaction-free measurements as
momentum-exchange free measurements Simon and Platzman derived
``fundamental limits'' on the IFM. They argued that the IFM can be
performed only on infinitely sensitive bomb and that a bomb which is
infinitely sensitive to any momentum transfer could not be placed at
the vicinity of the IFM device from the beginning. These arguments
fail because the EV IFM are not defined as momentum-exchange free
measurements. (Probably, the misconception came because of frequent
mentioning of Dickes' paper who concentrated on the issue of momentum
exchange in his procedure.) 

The arguments, similar to those of Simon and Platzman are relevant for
performing a modification of the EV IFM proposed by Penrose
\cite{Pen}. In the Penrose version of IFM the bomb plays the role of
one mirror of the interferometer. Thus, indeed, the uncertainty
principle put limits on placing the bomb in its place before the
experiment \cite{Pen-Vai}. In contrast, in the EV IFM the bomb need
not be localized prior to the measurement: the IFM  localizes it
by itself.

The ideal EV IFM need not be a momentum-exchange free experiment but
it might be.  If the object is localized before the IFM procedure,
then, indeed, the expectation value of momentum and of any power of
momentum of the object and that of the photon inside the
interferometer do not change during the time of the ``interaction''
between the photon and the object. (The time when the interaction
could take place or the time when the interaction took place in
another branch of the Universe.)  Such  procedures are usually considered as
momentum-exchange free. 

Aharonov \cite{Aha-pri} pointed out that the IFM process cannot take
place without exchange of any physical variable.  In the EV procedure
there is an exchange of {\em modular momentum}. The collapse of the
quantum wave of the photon from the superposition of the two wave
packets separated by a distance $a$ to a single wave packet continuing
to move in the upper part of the interferometer is accompanied by the
change in the momentum modulo $\hbar \over a$. Note that there is no
change in the  momentum modulo $\hbar \over a$ of the object (provided it was
localized at the lower arm of the interferometer from the beginning),
but, nevertheless,  the conservation law for the total modular
momentum is not contradicted because  the modulo momentum of the
object  is completely uncertain.

\vskip 1.cm \noindent
{\bf  8. The almost 100\% efficient  IFM.~~}
\vskip .2cm

In the IFM paper we have found a modification of the scheme presented
above which allows detection of almost 50\% of the bomb without
explosion (the rest explode in the process). At that point my belief
in the many-worlds interpretation lead us to a mistake: I persuaded Avshalom
Elitzur that we cannot do better. We wrote a footnote:
\begin{quotation}
  The MWI presents also a natural explanation why we cannot do
  better. Consider the world in which the photon hits the bomb. The
  world that replaces it in the case where the bomb is transparent
  interferes destructively with the world in which the detector $D_2$
  clicks. Since the latter is completely  eliminated it cannot have a
  probability larger than that of the former.
\end{quotation}
There is nothing wrong with the MWI. {\it I} made a mistake in the
framework of the MWI. I have not realized that one can devise an
experiment in which there are many different worlds in which the
photon hits the bomb (the hits take place at different times). All
these worlds should interfere destructively with the world in which
$D_2$ clicks. For this it is necessary that the sum of the amplitudes
of the worlds with the explosion will compensate the amplitude of the
world in which the bomb is detected without an explosion. If there is
a large number of ``explosion'' worlds, then the total measure of
existence of the worlds with explosion \cite{Va-mwi}, i.e. the
probability of explosion  can be arbitrary small even
when the sum of the amplitudes is large.

Our mistake was corrected by Kwiat at all. \cite{Kwi95}. They applied
quantum Zeno effect for constructing the IFM scheme which, in
principle, can be made arbitrary close to the 100\% efficiency. The
experiment with theoretical efficiency higher than 50\% have been
performed \cite{White98}.

Another claim about the IFM based on the reasoning in the framework of
the MWI \cite{Va-par}, I believe is correct. It is impossible to make
interaction-free measurement telling us that in certain place there is
no any object. Here, I mean ``interaction-free'' in the sense that no
photons (or other particles) pass through the place in question. The
argument is that our physical laws which include only local
interactions making  getting information about some
location without any particle being there paradoxical. In the case of the bomb,
the MWI solves the paradox by saying that since the laws apply to the
whole physical Universe which includes all the worlds, the reasoning
must be true only when we consider all the worlds. Since there are
worlds with the explosion we cannot say on the level of the physical
Universe that  no photons were at the location of the bomb. In
contrast, when there is no bomb, there are no other worlds. The
paradox in our world becomes the paradox for the whole Universe which
is a real paradox.

\vskip 1.cm \noindent
{\bf  9. Modifications of the EV IFM.~~}
\vskip .2cm

The almost 100\% efficient scheme of Kwiat et all. \cite{Kwi95} can be
described as follows. Two identical optical cavities coupled through a
highly reflective mirror. A single photon originally placed in the
left cavity. If the right cavity is empty, then after particular
number $N$ of reflections the photon  with certainty will be in the right
cavity. If, however, there is a bomb in the right cavity, the photon
with the probability close to 1 for large $N$  will be found in the left
cavity.  Testing at the appropriate time for the photon in the left
cavity, will tell us if there is a bomb in the right cavity.

This method keeps all conceptual features of the EV IFM. If the photon
is found in the left cavity, we are certain that there is an object in
the right cavity. If the object is an ultra-sensitive bomb or if it is
completely non-transparent object which does not reflect light
backwards (e.g., it is a mirror rotated by $45$ degrees relative to
the optical axes of the cavity as in the Kwiat et all. experiment)
then, when we detect the photon in the left cavity we can claim that
it never ``touched'' the object in the same sense as it is true in the
original EV method.

Another modification of the EV IFM which leads to the efficiency of
almost 100\% has been proposed by Paul and Pavicic \cite{PaPa} and
implemented in a laboratory by Tsegaye et all. \cite{Tse98}.  The
advantage of this proposal is that it has just one cavity, and
 is therefore easier to perform. The basic ingredient of this method is an
optical resonance cavity which is almost transparent when empty and is
an almost perfect mirror when there is an object inside.  However,
this method has a small conceptual drawback.  Always there is a
nonzero probability to reflect the photon even if the cavity is empty.
Thus, detecting reflected photon cannot ensure presence of the object
with 100\% certainty. This drawback has only academic significance. In
any real experiment there will be uncertainty anyway, and the
uncertainty which I mentioned can be always reduced below the level of
the experimental noise.

Other modifications of the IFM are related to interaction-free
``imaging''\cite{Kwi98} and interaction-free measurements of
semi-transparent objects \cite{Jang,Serge}.  These experiments hardly
pass the strict definition of the IFM in the sense that the photons do
not pass in the vicinity of the object. However, they all achieve a
very important practical goal, since we ``see'' the object reducing  very
significantly the irradiation of the object: this can allow measurements
on fragile objects.

\vskip 1.cm \noindent {\bf 10. Conclusions.~~} \vskip .2cm

I have reviewed various analyses, proposals, and experiments which appeared
following the method for the interaction-free measurement of Elitzur and
Vaidman. The common feature of all these proposals is that we obtain
information about an  object while significantly reducing irradiation
of the objects. The meaning of the EV IFM is that if an object changes
its 
 internal state (not the quantum state of its center of mass)
due to the radiation, then the method allows detection of the location
of the object without {\it any} change in its internal state. The IFM
allow measurements of position of infinitely fragile objects. In some
sense it locates objects without ``touching'', i.e. without particles
of any kind passing through its vicinity. Contrary to recent claims,
such IFM have no any fundamental limit.

We should mention that the interaction-free measurements do not have
vanishing interaction Hamiltonian. In general, the IFM are also not
energy-exchange free or momentum-exchange free processes: the IFM can
change very significantly the quantum state of the observed object and
we still name it interaction-free. On the other hand the method do
allow performing some non-demolition measurements. It might be
momentum-exchange free and energy-exchange free.

 Numerous papers on the IFM interpreted the concept of
 interaction-free in many different ways. I hope that in this work I
 clarified the differences and stated unambiguously the meaning of the
 original proposal.

\vspace{.3cm}
 \centerline{\bf  ACKNOWLEDGMENTS}
 
 It is a pleasure to thank Yakir Aharonov, Gideon Alexander and Philip
 Pearle for helpful discussions. 
 This research was supported in part by grant
 471/98 of the Basic Research Foundation (administered by the Israel
 Academy of Sciences and Humanities).

\end{document}